# RRL: A Rich Representation Language for the Description of Agent Behaviour in NECA

Paul Piwek
ITRI, Faculty of Science and
Engineering, University of Brighton
Watts Building, Moulsecoomb
BN2 4GJ Brighton, UK
+44 (0) 1273 64 29 16

Paul.Piwek@itri.bton.ac.uk

Brigitte Krenn
ÖFAI, Austrian Institute for Artificial
Intelligence
Schottengasse 3,
A-1010 Vienna, Austria
(+43-1) 5324621-2

brigitte@ai.univie.ac.at

Marc Schröder
DFKI, German Research Center for
Artificial Intelligence
Stuhlsatzenhausweg 3
D-66123 Saarbrücken, Germany
+49 (0) 681-302-5303

Marc.Schroeder@dfki.de

Martine Grice
IPUS, Institute for Phonetics,
University of the Saarland
Building 17.2, Postfach 151150
66041 Saarbrücken
+49 (0) 681 302-4696

mgrice@coli.uni-sb.de

Stefan Baumann
IPUS, Institute for Phonetics,
University of the Saarland
Building 17.2, Postfach 151150
66041 Saarbrücken
+49 (0) 681 302-4244

baumann@coli.uni-sb.de

Hannes Pirker
ÖFAI, Austrian Institute for Artificial
Intelligence
Schottengasse 3,
A-1010 Vienna, Austria
(+43-1) 5324621-3

hannes@ai.univie.ac.at

## ABSTRACT
In this paper, we describe the Rich Representation Language (RRL) which is used in the NECA system. The NECA system generates interactions between two or more animated characters. The RRL is a formal framework for representing the information that is exchanged at the interfaces between the various NECA system modules.

## 1. INTRODUCTION
The objective of the work presented here is the development of a rich formal language suitable for representing the behaviour of conversational agents. The work is being carried out within the NECA project (a Net Environment for Embodied Emotional Conversational Agents; see http://www.ai.univie.ac.at/NECA/). The project focuses on communication between animated characters that exhibit credible personality traits and affective behaviour. Two concrete application scenarios are being developed. The first scenario, 'eShowRoom' concerns car sales dialogues between a seller character and one or more buyer characters. The second scenario 'Socialite' uses the same technology for a different purpose: users create their avatars and send them into a virtual world. The users can then watch animated scenes featuring their avatars in interaction.

The NECA system generates the interaction between two or more characters in a number of steps, with the information flow proceeding from a Scene Generator to a Multi-modal Natural Language Generator, to a Speech Synthesis component, to a Gesture Assignment component, and finally to a media player. Thus a representation language is needed in NECA as a means for representing the various kinds of expert knowledge required at the different interfaces between the components in the NECA architecture. This paper discusses the information required at the interfaces between the NECA system modules. We call the formal framework for representing this information the Rich Representation Language (RRL). This paper aims at introducing the RRL at the conceptual level.

## 2. REQUIREMENTS FOR RRL
As in any area of software development, the specification of the RRL should be based on an analysis of the requirements of its use. We discern three main sources for the requirements for the RRL.

**I**. **The applications domains**. **A**. *Ability to represent combinations of different types of information*. Our application domains involve representations of scenes with two or more characters. The realization of such scenes requires the representation of various types of information such as semantic content, pragmatic force, morpho-syntax, information structure, graphemic and phonemic form, prosody, facial expression, posture and gesture. Because information of one type often 'points into' information of another type the RRL should be able to represent combinations of different types of information. **B**. *Expressive power required by the domain*. Different types of information require different levels of expressive power. For instance, semantic representations can usually be expressed in terms of context-free grammars whereas the simultaneous representation of syntactic structure and information structure seems to require overlapping tree structures.

**II**. **The operation of the modules that manipulate the representation**. **A**. *Ease of manipulation*. Algorithms for changing the representations should be easy to write. Ideally, incremental construction of the representations should be supported. **B**. *Ease of search*. The representations should allow for simple and robust algorithms to search for items. For instance, in the context of spoken dialogue systems it has been argued that flat representations are particularly suited for these purposed (see e.g., [14]). On the other hand, for document representation tree structures have become the standard solution.

**III**. **The developers of the system who implement and maintain algorithms and resources for manipulating the representations**. Maintainability requirements apply on two levels: 1. on the level of actual RRL representations which the system manipulates and 2. on the level of their specification, e.g., by means of XML schema or a description logical T-box. It is very well possible that satisfying a requirement on one of the levels is incompatible with satisfying it on the other level. **A**. *Predictability*. Once one bit of the representation/specification has been understood, it should be easy to predict what other parts will look like. **B**. *Locality*. Changes to one part of a representation/specification should not affect other parts, i.e., the representation/specification should be as modular as possible. (E.g., the introduction of a new attribute of an object of type *A* should not affect objects of other types). **C**. *Conciseness*. Short representations/specifications are preferred over lengthy ones. **D**. *Intelligibility*. Wherever possible informative names and abbreviations should be used. Furthermore, graphical representations are often more transparent to humans. Therefore a formalism that allows for a natural graphical depiction might be favoured.

## 3. SCENE DESCRIPTIONS

Following terminology that is common from the theatre[1], we call an interaction between our synthetic characters a scene (see also [1]'s notion of a presentation team). In this section we discuss the RRL representations of scenes that are constructed by the Scene Generator and then sent to the Multimodal Natural Language Generator of the NECA system. A Scene Description specifies the content, type, temporal order and emotion of the acts that the characters engage in. To compute the emotions of the acts (type, intensity and cause), the scene generator incorporates a dedicated module, the so-called affective reasoner (see [12]).

In NECA, the XML standard is used throughout the project to represent the syntax of the RRL. However, the XML representations are derived from a more abstract object-oriented network representation format. This abstract level of representation allows us to think about the information that needs to be encoded in terms of objects and their attributes without worrying about the linear/hierarchical representation of this information in XML. Of course, at some point we do need a systematic mapping of these networks to XML (see http://www.ai.univie.ac.at/NECA/RRL/ for examples of the XML encoding of the representation which are proposed here), but we argue that at the content design stage it is better to abstract over that issue.

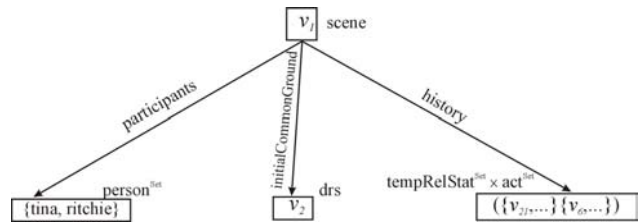

**Figure 1 Graphical depiction of a network representation of a scene**

A network is a collection of labelled nodes that are connected by labelled arrows (directed arcs). Nodes stand for simple or complex objects. We discern three types of nodes: (**1**) Nodes that stand for arbitrary objects. Such a node conveys the existence of an object, but does not refer to a specific object. We call a node of this sort a variable. (**2**) Nodes that stand for specific named objects. Nodes of this sort are called constants. (**3**) Complex Nodes. An example of a complex node is, for instance, a node that represents a *set* or *list* of nodes of type (1) or (2).

Nodes are labelled. The label of a node conveys the *type* of the object(s) that the node represents. Finally, nodes can be connected to each other by means of labelled arrows. If one node, say $v_1$, is connected to a second node, say $v_2$, by means of an arrow labelled $Attr_1$, we interpret this as meaning that the *value* of the *attribute* $Attr_1$ of node $v_1$ is equal to $v_2$. For a complete formal specification we refer to [16]. There it is shown how Scene Descriptions can be specified for one or more domains using a T-box in a fairly transparent and concise manner (cf. requirements **III**.**C** and **D**) and bring the benefit of being encodable as flat representations (cf. requirement **II**.**B**).

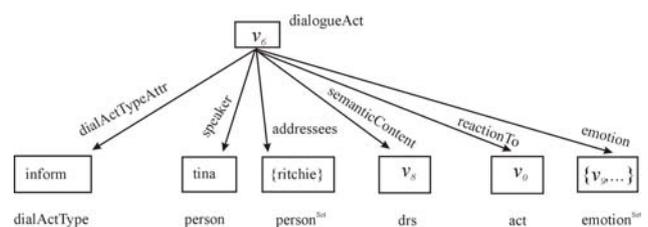

**Figure 2 Graphical depiction of part of a network representation of a dialogue act**

Now, let us consider some examples of network representations for Scene Descriptions taken from the eShowRoom car sales domain (see http://www.ai.univie.ac.at/NECA/RRL/ for a screen dump of the setting of such a scene). At the root of such a network (see Figure 1; this Figure only shows the upper part of the network) there is a node ($v_1$) representing the scene itself. The scene has three attributes with the following values: 1. the set of persons/characters who are participants in the scene, 2. a Discourse Representation Structure (DRS; [11]) which represents

---

[1] "Scene **I** Theatr. **1** A subdivision of (an act of) a play, in which the time is continuous and the setting fixed, marked in classic drama by the entrance or departure of one or more actors and in non-classic drama often by a change of setting; the action and dialogue comprised in any one of these subdivisions." (source: Electronic New Shorter Oxford English Dictionary, 1996)

the common ground amongst the interlocutors at the start of the conversation, and 3. a history which consists of a set of temporal relation statements and a set of acts.

Scene Descriptions are the input of the Multimodal Natural Language Generator (M-NLG). The M-NLG determines the form (both linguistic and non-linguistic) of the acts that are part of the scene. In order to determine the linguistic form, the (initial) common ground (e.g., [4]) of the interlocutors is indispensable. For instance, whether an object is referred to with definite noun phrases depends on whether it is a part of the common ground. The common ground also plays an important role in the selection of the content of such expressions (cf. [7]). For the formal representation of the common ground we borrow the notion of a DRS from Discourse Representation Theory (DRT). We provide more details on the use of DRT in the RRL later in this section.

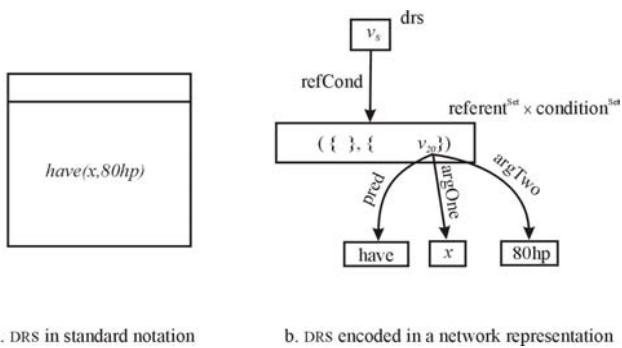

a. DRS in standard notation    b. DRS encoded in a network representation

**Figure 3**

The history of a Scene Description consists of two ingredients: the set of acts that are part of the scene and a specification of their temporal ordering. The two are represented separately: each act has a unique identifier (a variable); the ordering on the acts ($v_1$ is before $v_2$, $v_1$ and $v_2$ are simultaneous, etc.) is specified on the identifiers. Our main motivation for this separation is that we want to allow for underspecification of the temporal ordering. Sometimes it is preferable to defer decisions on ordering to the M-NLG where the precise phrasing and gestures with which a dialogue act is realized are known.[2] (Underspecification falls under requirement **I.B**). Let us now proceed to discuss the representation of individual dialogue acts.[3]

---

[2] For example, suppose one character asks the question: 'Would you like to know more about the motor of this car?'. The Scene Description might say that this act is followed by a positive feedback act and the question 'How much horse power does it have?'. These might be realized in *sequence* as 'Yes, how much horse power does it have?' or *simultaneously*, as 'How much horse power does it have?', where the positive feedback is implicit in the question. Our aim is to have a representation format that is compatible with both realizations. It therefore requires the means for expressing underspecification of temporal relations between dialogue acts.

[3] In our model, 'dialogue acts' are a subtype of the type 'act'. We discern a second subtype of 'act' called 'non-communicative act'. These encompass for example the act of a character moving/walking from one location to another.

We work backwards by investigating the question what type of representation would be needed to generate: "This wonderful car has 80 hp". Figure 2 shows part of such a representation. It specifies the dialogue act type, the speaker, the set of addressees, the semantic content, the dialogue act to which it is a response (e.g., a question) and the emotion.

Here we focus on semantic content and emotion. To represent the semantic content we use the DRSs of DRT. DRT was originally developed as a theory of discourse interpretation. It has been applied successfully to a wide range of discourse phenomena such as presupposition, anaphora, plurality, tense and information structure. It provides a rich and well-tested basis for expressing semantic content. There is, however, a complication. We need to somehow integrate DRSs into our network representations. Fortunately, this can be achieved in a fairly straightforward manner by reifying DRSs and their constituents as objects in their own right. This leads to a uniform representation of various types of dialogue content within a single network (cf. requirement **I.A.**). Additionally, it provides us with handles for DRSs and their constituents. These are not available in standard DRT representations. Below an example is given of how such handles can be put to use when DRSs are integrated with other types of dialogue information.

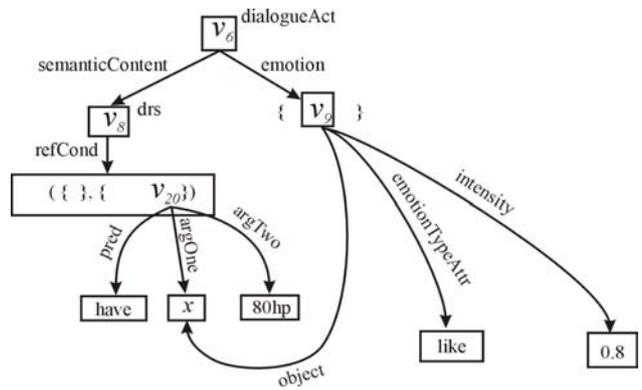

**Figure 4 Graphical depiction of an integrated/uniform representation of semantic content and emotion**

Figure 3.a contains a DRS in the standard notation. It is a box consisting of two parts: a set of referents (which is empty in this particular case) and a set of conditions. In Figure 3.b the corresponding network representation is depicted. The DRS, the referents and the conditions are represented as objects in the network. In particular, the condition *have(x,80hp)* is expressed as the object $v_{20}$ which has attributes representing its predicate 'have' and its two arguments: argOne with the value $x$ and argTwo with the value 80hp. Figure 3.b encodes a DRS which can be paraphrased as 'The car has 80hp'.[4]

---

[4] The car corresponds to the object $x$. It is not introduced as a referent in the DRS, as we assume that it is already part of the common ground. This licenses the use of a definite noun phrase (of course, it also has to be the most salient car in the common ground or be uniquely identifiable in the common ground to warrant the use of a definite).

In order to enable the agents to express emotions, the emotion first needs to be determined, and subsequently expressed through wording, prosody, facial expression, and body language. Research traditions dealing with different parts of this process have different paradigms for the description of emotion ([5]). In affective reasoning, an appraisal-based framework is used ([15]; henceforth OCC), mapping emotion-eliciting conditions onto a large number of emotion categories; in speech synthesis, recent research suggests it may be beneficial to represent emotional states by means of emotion dimensions, i.e., level of arousal, evaluation/valence, and possibly dominance ([6]; [18]); and facial expression research has traditionally described emotions using a small number of basic emotion categories ([8]).

As the NECA system encompasses all of these aspects, we try to accommodate all of them by using a complementary representation. First, the affective reasoner (which is part of the Scene Generator; see [12]) determines the emotion category, its intensity and possibly the object that caused it. Consider Figure 4 (which omits attributes which are not relevant to the current discussion). There the benefit of uniform network representations is illustrated. $v_9$ represents the emotion with which the dialogue act $v_6$ should be expressed. It not only specifies the type and intensity (in terms of the OCC model) of the emotion, but also indicates the cause (again following OCC). A pointer into the semantic representation is used to identify the cause. For the wording of the dialogue act this could, for instance, lead to the inclusion of an evaluative adjective as in: 'The wonderful car has 80hp'.

For speech the OCC emotion category is mapped onto emotion dimensions, using co-ordinate values from the literature ([24]). For facial expression, the OCC emotion categories generated by the affective reasoner are mapped onto their closest basic emotion category, or onto an interpolated facial expression ([22]).

## 4. SPEECH SYNTHESIS

The task of the speech synthesis and gesture assignment components is to realise the output of the M-NLG through voice and animation. The speech synthesis component needs to process the data first, as it generates timing information required by the Gesture Assignment module (see Section 5.2). The speech synthesis interprets the natural language text and additional information about its meaning and adds prosodic structure. This comprises information on prosodic phrasing and accentuation as well as (emotionally adequate) intonation contours. The enriched document, along with a generated audio file, is passed on to the Gesture Assignment module.

The task for speech synthesis is to convey, through adequate voice quality and prosody, the intended meaning of the text as well as the emotion with which it is uttered. In text-to-speech systems, the realisation of the words and their prosody is based on a shallow local syntactic analysis of the text. No semantic information can usually be inferred. However, global syntactic structure as well as semantic information and information structure are known to be important factors determining the intonation ([13], [21]). In order to obtain the best quality of synthetic speech, it is therefore desirable to represent these types of information in the input to the speech synthesis system, making this component more similar to a concept-to-speech system (e.g. [10]) than to a text-to-speech system. However, it should also be possible to specify orthographic input by hand, in which case no linguistic annotation is available. Combined, this requirement amounts to a scalable representation language, which should be able to contain – alongside paralinguistic information (e.g., emotions) – in the minimal case little more than the words to be spoken, and in the maximal case, a full specification of linguistic structure.

In the minimal case, the M-NLG (or a human developer) only provides text with no linguistic annotation to the speech synthesis component, along with information about the speaker of a dialogue act and the emotion with which it is to be spoken. The speech synthesis determines the appropriate prosody and voice quality based on the specified emotion dimensions ([18]). The text-to-speech part of the synthesis component utilizes the sparser representation to drive the text-to-speech rules.

In the maximal case, the M-NLG additionally provides detailed linguistic information about the text structure. This includes the part-of-speech for each word, and optionally a phonetic transcription for irregular words not in the lexicon. The syntactic structure is fully specified, providing the full syntactic tree of phrase nodes and their grammatical functions. In addition, the information structure (in terms of theme and rheme) as well as the informational status of individual referents (in terms of givenness and contrast) is specified.

A challenge for the RRL is the simultaneous specification of syntactic structure, information structure and prosodic structure, since there is the possibility of overlap ([21]), corresponding to crossing edges in the respective tree structures (cf. requirements **I.A** and **B**). Crossing edges are not permitted in XML, which requires a strictly embedding tree structure. Therefore, when two or more potentially overlapping structures are to be encoded in the same tree, only a subset of the occurring phenomena can be fully described. There are at least two possible strategies to overcome this problem: 1) A "flat" encoding of one of the conflicting structures, using XML empty elements, thus circumventing the XML structure limitations; 2) A separation of conflicting structures into several autonomous hierarchies which are linked through a reference mechanism. As for the encoding of prosodic structure (i.e. intonation phrases), the use of a "flat" representation (i.e. solution 1) is feasible, since it reflects one of the current theoretical approaches to prosody ([20]) and is encodable using the tonal annotation system ToBI (e.g. [2]). The German adaptation of ToBI (Grice et al, to appear) is already part of NECA's TTS system MARY ([19]). As for the possible conflicts between syntax and information-structure, the current draft of the NECA RRL specification (http://www.ai.univie.ac.at/NECA/RRL/) assumes that these structures can still be represented within a single XML hierarchy by only applying slight restrictions on the possible configurations for information structure.

## 5. GESTURE ASSIGNMENT

Gesture assignment in the NECA architecture is distributed over three levels in the information flow during processing. The term gesture here is used in a broad sense subsuming facial expression, gesture proper and posture.

- Assignment of candidate gestures related to the scene to be generated, semantic content of utterances, turn-taking etc. takes place within the M-NLG.

- Elaboration of gesture assignment after speech synthesis, i.e. specification of the exact timing and adding "low level"-physiological gestures such as eye-blinking and breathing. This takes place in the Gesture Assignment module.

- Transformation of the gestural specifications from RRL to player-specific formats, such as MPEG-4 FAPs. This is not part of the RRL proper and is therefore not addressed in this paper.

## 5.1 Gesture Assignment within Multimodal Natural Language Generation (M-NLG)

Within the NECA architecture, the task of M-NLG is to determine integrated linguistic and non-linguistic behaviour. On the one hand it generates text richly annotated with information on syntax, information structure and emotion which will feed into the speech synthesis component. On the other hand it determines non-linguistic aspects of character behaviour such as iconic and emblematic gestures, a number of facial expressions[5], and emotive interjections (also referred to as affect bursts).[6] As mentioned in section 4 information on the exact timing of utterances, syllables and phonemes is still lacking at this stage. Because timing information is indispensable, the final specification of non-linguistic acts has naturally to be postponed. In our architecture the M-NLG thus has the role of mainly providing gesture candidates, while the selection, final specification and scheduling is performed later on. (See the BEAT-system for a related approach; [3])

In the M-NLG, gesture information can already be assigned at different levels of structure, e.g. whole dialogue acts (as may be the case for a certain posture), or at phrase or word level. The following classes of gestures are available at this stage:

- *emblematic gesture* (gestures which are conventionalized such as yes/no);

- *iconic gesture* (mimicking the form of an object or action, such as imitating a telephone receiver by stretching thumb and little finger between ear and mouth);

- *deictic gesture* (pointing gestures with arm and hand);

- *contrast gesture* (usually literally a "on the one hand… on the other hand" gesture);

- *turn accompanying gesture, posture and facial expression* (e.g., eyebrow raise, eye gaze);

- *emotional nonverbal expression: facial expression, posture* (e.g., "hanging shoulders" signalling sadness);

- *back channelling gestures* (e.g., nodding, frowning but also such gestures combined with interjections such as "aha").

Gesture attributes are priority (aiding behaviour selection), intensity, direction, and stretch/size.

## 5.2 Gesture Assignment after Speech Synthesis

After speech synthesis and in addition to the gestures assigned by M-NLG, the final Gesture Assignment module (GA) is provided with the following information to be used for fine-tuning the non-linguistic acts and for synchronising speech and gesture. Information derived throughout the speech-synthesis process, in particular results of prosody determination are made visible to the GA-component.[7]

- *phones*: the name and exact temporal position of each speech sound is provided. Apart from being used for specifying the visemes for lip-synchronous animation, the timing of the single sounds provides the temporal backbone from which the timing of the other elements can easily be calculated;

- *syllable and word-boundaries* (e.g., eye movements are tightly synchronised with syllables, beats synchronise with emphasized words);

- *syllables bearing word stress* (stressed syllables are the preferred anchor point for e.g. deictic gestures, eyebrow raising, head nods, eye blinking);

- *position and type of sentence accents* (i.e., stroke gestures preferentially coincide with syllables bearing a pitch accent);

- *prosodic phrases* (i.e., position and type of intonational boundary tones; prosodic phrases act as a landmark for eyebrow raising, head nods, and eye blinking);

- *pauses* (e.g., used for the timing of posture changes, breath movements, head nods).

---

[5] Including eye gaze, which can have a number of functions in communication (see, e.g., [17]).

[6] Emotive interjections, such as sigh, yawn, laughing etc. add to the emotional believability of the agents' utterances, and are produced holistically (as opposed to being generated from smaller units in the speech synthesis). Technically speaking this can be handled by including tags specifying interjections within the text to be spoken, as no overlap/parallelism with speech can occur.

[7] For examples on the relation between speech parameters and gesture, and further literature on the topic see [23].

As a final step in gesture assignment, physiologically based animations are scheduled in accordance with the constraints imposed by the content based animations. For example, *physiological eye blinking* (to make figures look more natural we introduce a facultative blink beat) and *physiological breathing* (regular rib cage movements are assigned to make the animation of the characters look more natural).

## 6. DISCUSSION

The RRL is a special purpose markup language which represents a wide range of expert knowledge required at the interfaces between the different components in the NECA architecture. Existing markup languages have either been designed for the representation of information at individual levels of description or provide a combination of markups at different levels of representation for multimedia annotation. A good deal of work has been done on the former, especially on speech synthesis markup and facial animation coding, see, e.g., the W3C Speech Synthesis Markup Language[8] and MPEG4[9] FAPs (Facial Animation Parameters). For a recent survey of facial and gesture coding schemes see the ISLE Report D9.1. ([23]). Markup languages for multimedia annotation include VHML[10], SMIL[11], MPML[12] and TVML[13]. Especially VHML aims at unifying a confederation of existing special purpose markup languages.

The RRL differs from other multimedia markup languages in that these are typically designed to support a fairly text-based annotation of multimodal input to media players, ideally in a rather generalized and standardized way, whereas the RRL is in addition capable of representing expert knowledge which may be created by a processing component rather than a human author (for instance, detailed information on the linguistic structure). In developing the RRL we are able to draw on existing standardisation efforts and build on well-defined cores of XML-based markup languages, especially in the field of speech synthesis and facial animation. On the other hand, experience gained from our efforts to represent expert markup at all levels of representation (from rather abstract representations of scenes to more or less player specific representations which determine the final output of the NECA system) will hopefully feed into future standardisation efforts.

## 7. ACKNOWLEDGMENTS

This research is supported by the EC Project NECA IST-2000-28580. The information in this document is provided as is and no guarantee or warranty is given that the information is fit for any particular purpose. The user thereof uses the information at its sole risk and liability. We would like to thank our colleagues in the NECA project, in particular, Erich Gstrein, Bernhard Herzog, Martin Klesen, Thomas Rist and Kees van Deemter for discussions and comments on this paper and for their help with the preparation of the supplementary material for this paper which can be found at http://www.ai.univie.ac.at/NECA/RRL/.

---

[8] http://www.w3.org/TR/speech-synthesis/

[9] http://www.es.com/mpeg4-snhc/index.html

[10] http://www.vhml.org/

[11] http://www.w3org/AudioVideo/

[12] http://www.miv.t.u-tokyo.ac.jp/HomePageEng.html

[13] http://www.strl.nhk.or.jp/TVML/index.html